\documentclass[10pt,aps,prl,twocolumn,groupedaddress,longbibliography,
showpacs]{revtex4-1}
\usepackage{epsfig,amsmath,amssymb,MnSymbol,verbatim,bbm}
\usepackage[colorlinks=true,citecolor=blue]{hyperref}

\begin{document}


\title{Thermodynamic Geometry of Microscopic Heat Engines}
\author{Kay Brandner\textsuperscript{1}}
\author{Keiji Saito\textsuperscript{2}}
\affiliation{\textsuperscript{1}Department of Applied Physics,
Aalto University, 00076 Aalto, Finland}
\affiliation{\textsuperscript{2}Department of Physics, 
Keio University, 3-14-1 Hiyoshi, Yokohama 223-8522, Japan}




\begin{abstract}
We develop a geometric framework to describe the thermodynamics of 
microscopic heat engines driven by slow periodic temperature 
variations and modulations of a mechanical control parameter. 
Covering both the classical and the quantum regime, our approach 
reveals a universal trade-off relation between efficiency and power 
that follows solely from geometric arguments and holds for any 
thermodynamically consistent microdynamics. 
Focusing on Lindblad dynamics, we derive a second bound showing
that coherence as a genuine quantum effect inevitably reduces the
performance of slow engine cycles regardless of the driving 
amplitudes. 
To demonstrate the practical applicability of our results, we work out
the example of a single-qubit heat engine, which lies within the range
of current solid-state technologies. 
\end{abstract}


\maketitle

\renewcommand{\H}{\mathsf{H}}
\renewcommand{\S}{\mathcal{S}}
\newcommand{\D}{\mathsf{D}}
\newcommand{\T}{\tau}
\newcommand{\F}{\mathcal{F}}
\newcommand{\R}{\mathcal{R}}
\newcommand{\W}{\mathcal{W}}
\newcommand{\A}{\mathcal{A}}
\renewcommand{\L}{\Lambda}
\newcommand{\LL}{\mathcal{L}}
\renewcommand{\r}{\rho}
\newcommand{\ce}{\varepsilon}
\renewcommand{\l}{\lambda}
\newcommand{\lb}{\boldsymbol{\l}}
\newcommand{\Lb}{\boldsymbol{\L}}

\newcommand{\tr}[1]{{{{\rm Tr}}}\bigl[ #1 \bigr]}
\newcommand{\av}[1]{\langle #1 \rangle}

\newcommand{\ixt}{\raisebox{-1.6pt}{{{\scriptsize $t$}}}}
\newcommand{\ix}[1]{\raisebox{-1.6pt}{{{\scriptsize $#1$}}}}

\renewcommand{\tint}{\int_0^\T \!\!\!\! dt \;}
\newcommand{\dbar}{{\mkern3mu\mathchar'26\mkern-12mu d}}

\vbadness=10000
\hbadness=10000

The laws of thermodynamics put fundamental limits on the performance
of thermal machines across all length and energy scales. 
A prime example is the Carnot bound on efficiency, which applies to 
James Watt's steam engine as well as to recent small-scale engines 
using colloidal particles \cite{Blickle2011,Martinez2015a,
Martinez2015}, single atoms \cite{Abah2012,Rossnagel2015} or 
engineered quantum systems \cite{Klatzow2017a,Ronzani2018}. 
Still, despite its universality, this bound is mostly of theoretical 
value as it can be attained only by infinitely slow cycles producing 
zero power. 
Practical devices, however, must operate in finite time and therefore
are inevitably subject to frictional energy losses suppressing their 
efficiency. 
Hence, we are prompted to ask: how much performance has to be 
sacrificed for finite speed?

This question, which inspired the development of finite-time 
thermodynamics in the 1970s \cite{Andresen2011}, has recently 
attracted renewed interest:
triggered by the observation that Carnot efficiency at finite power 
could indeed be possible in systems with broken time-reversal 
symmetry \cite{Benenti2011}, a series of studies discovered 
quantitative trade-off relations, which limit the finite-time 
performance  of microscopic engines and confirm the conventional
expectation that dissipationless machines deliver zero power
\cite{Brandner2015f,Brandner2015,Proesmans2015a,
Proesmans2016c,Brandner2016,Shiraishi2016b,Pietzonka2017,
Shiraishi2018b}. 

\begin{figure}
\includegraphics[scale=.45]{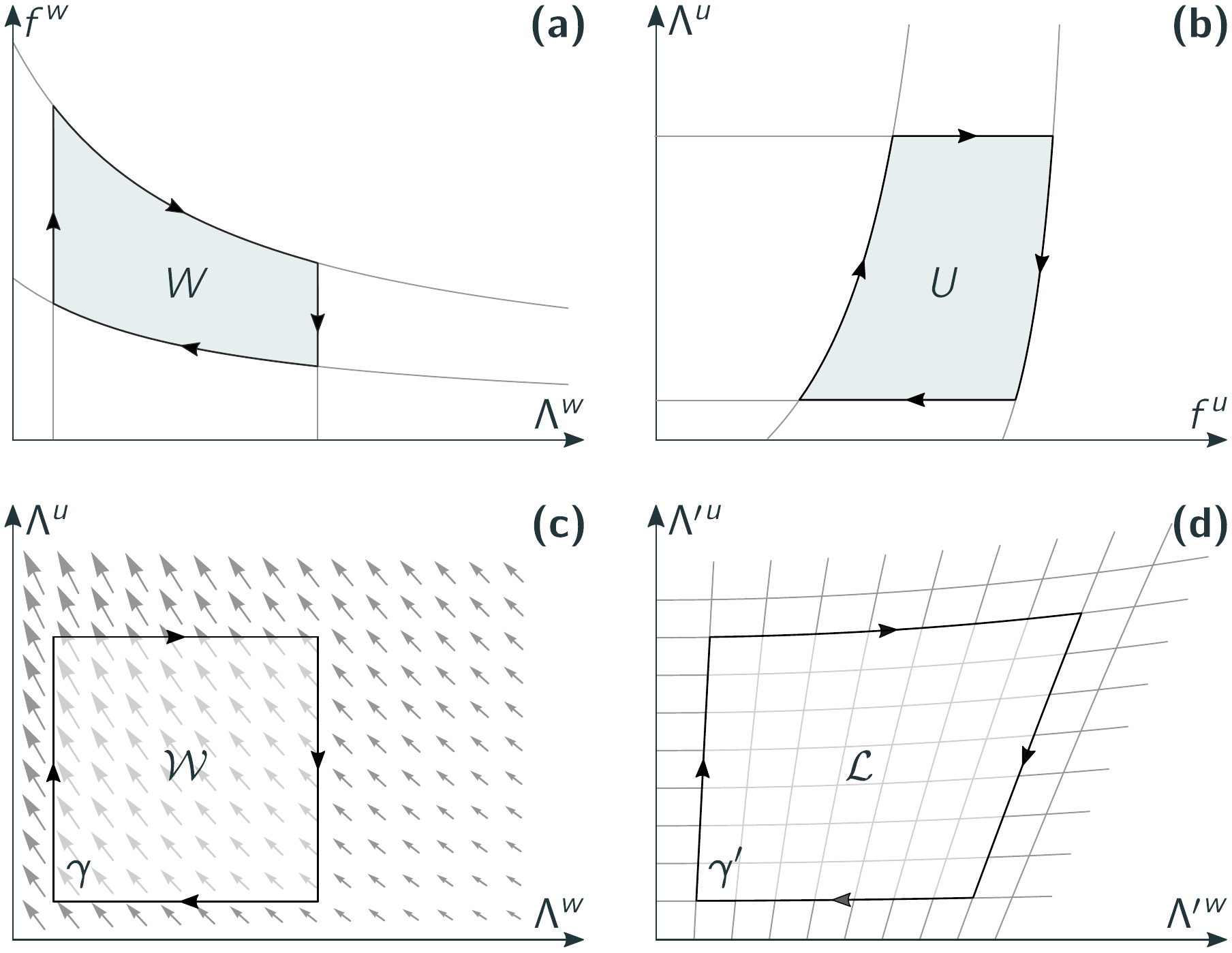}
\caption{Four faces of a microscopic engine cycle. 
Upper panel: Thermodynamic picture. 
The two sketches show effective pressure-volume (a) and 
temperature-entropy (b) diagrams for a Stirling cycle consisting of 
two \emph{isochoric} ($\L^w={{{\rm const}}}$) and two 
\emph{isothermal} ($\L^u={{{\rm const}}}$) strokes. 
The enclosed areas correspond to the generated work $W$ and
the effective thermal energy uptake $U$.
Lower panel: Geometric picture. 
In the space of control parameters $\L^u$ and $\L^w$, the 
adiabatic work $\W$ is given by the line integral of the thermodynamic
vector potential along $\gamma$, i.e., the flux of the corresponding
effective magnetic field through the area encircled by this path (c).
In the curvilinear coordinates $\L'^w$ and $\L'^u$, which carry the 
thermodynamic metric, $\gamma$ is distorted into the contour 
$\gamma'$, whose length $\LL$ provides a lower bound on the 
dissipated energy (d). \label{FigStirling}}
\vspace*{-.5cm}
\end{figure}

These results rely on stochastic models to describe the internal 
dynamics of small-scale engines. 
Here, we pursue an alternative strategy that builds on the framework 
of thermodynamic geometry \cite{Ruppeiner1995}. 
This approach replaces the traditional thermodynamic picture, which 
mixes control and response variables, with a geometric picture. 
The properties of the working system are thereby encoded in a vector 
potential and a Riemannian metric in the space of control parameters,
see Fig.~\ref{FigStirling}. 
The driving protocols define a closed path in this space and can
thus be assigned an effective flux and length. 
In adiabatic response, these quantities provide measures for the two 
key figures of merit: the work output and the minimal dissipation
of the underlying thermodynamic process. 

The idea of using geometric concepts to describe the thermodynamics of
finite-time operations was originally conceived for macroscopic 
systems and developed mainly on the basis of phenomenological 
principles \cite{Berry1984a,Ruppeiner1995,Andresen2011}.
Over the last decades, this approach has been formulated on microscopic
grounds \cite{Brody1995}, linked to information-theoretic quantities
\cite{Ruppeiner1995} and extended to classical nano-scale systems
\cite{Crooks2007}, closed quantum systems far from equilibrium 
\cite{Deffner2013b} and, most recently, open quantum systems 
\cite{Scandi}. 
Thermodynamic geometry has thus become a powerful tool, which, as its 
key application, provides an elegant way to determine optimal 
control protocols minimizing the dissipation of isothermal processes
\cite{Zulkowski2012,Sivak2012,Rotskoff2015,Machta2015,Scandi}. 
Yet, this framework has neither been applied systematically to bound 
the performance of cyclic micro-engines nor to explore the impact of
quantum effects on such devices.  

To progress in this direction, we consider a general model for a 
microscopic heat engine consisting of two components: 
a working medium with tunable Hamiltonian $H_\l$ and a heat source to
control the temperature $T$ of the environment of this system. 
The device is operated by periodically changing the parameters $\l$
and $T$ such that the vector
\begin{equation}\label{SPara}
\Lb\equiv (T,\l) \equiv (\L^u,\L^w)
\end{equation}
passes through a closed path $\gamma:t\mapsto\Lb_t$. 
Once the system has settled to a periodic state $\rho_t$, the average 
output and input of this process, i.e., the mean generated work and 
the effective uptake of thermal energy from the heat source, are given
by
\begin{equation}\label{SDefUW}
W = \tint f^w_t \dot{\L}^w_t  \quad\text{and}\quad
	U = -\tint f^u_t \dot{\L}^u_t. 
\end{equation}
Here, dots indicate time derivatives, $\T$ denotes the cycle time and 
the thermodynamic forces,
\begin{equation}\label{SThDForces}
f^w_t\equiv -\tr{\rho_t\partial_{\l} H_{\l}}\bigl|_{\l=\l\ix{t}}
	\quad\text{and}\quad
	f^u_t\equiv -\tr{\rho_t\ln\rho_t},
\end{equation}
correspond to the generalized pressure and the entropy of the working 
system, respectively.  
The upper panel of Fig.~\ref{FigStirling} shows a graphical
illustration of this scheme. 
Note that we set Boltzmann's constant to $1$ throughout.

Using the relations \eqref{SDefUW}, the energy balance of the engine
can be formulated as 
\begin{equation}\label{SDissEnergy}
A\equiv U-W = - \tint f^\mu_t\dot{\L}^\mu_t \geq 0, 
\end{equation}
where $\mu=w,u$ and summation over identical indices is understood
throughout. 
The quantity $A$ corresponds to the mean energy loss or 
\emph{dissipated availability} \cite{Salamon1983} per cycle; it must 
be non-negative as a direct consequence of the second law
\footnote{To this end, observe that the first law, $\dot{E}_t=
J_t-P_t$, implies $-f^\mu_t \dot{\L}^\mu_t = T_t\Sigma_t
+\partial_t(E_t - T_t S_t)$ and that \mbox{$\Sigma_t\equiv
\dot{S}_t-J_t/T_t\geq 0$} corresponds to the total rate of entropy 
production. 
Here, $E_t$ and $S_t= f^u_t$ denote the internal energy and entropy of
the system, respectively, $J_t$ is the rate of heat uptake
from the environment and $P_t$ the instantaneous power. 
}. 
Thus, the dimensionless coefficient 
\begin{equation}\label{SEfficiency}
\eta\equiv W/U\leq 1
\end{equation}
provides a proper measure for the efficiency of the engine.
This figure is well-defined for any control protocols leading to 
positive work extraction, i.e., $W> 0$. 
In the special case, where the temperature switches between two 
constant levels,  it becomes an upper bound on the traditional
thermodynamic efficiency, which involves only the heat uptake during
the hot phase of the cycle \footnote{
Specifically, we have 
$\eta_{{{\rm th}}}/\eta_{{{\rm C}}}\equiv W/Q\eta_{{{\rm C}}}\leq\eta$,
where $Q$ denotes the heat uptake during the hot phase, 
$\eta_{{{\rm C}}}\equiv1-T_{{{\rm c}}}/T_{{{\rm h}}}$ is the
Carnot factor and $T_{{{\rm h}}}$ and $T_{{{\rm c}}}<T_{{{\rm h}}}$
correspond to the hot and the cold temperature.}.

Under quasi-static driving, the system follows its instantaneous 
equilibrium state, i.e., we have 
\begin{equation}\label{SAdState}
\rho_t = \varrho_{\Lb\ix{t}}
	\quad\text{with}\quad
	\varrho_{\Lb}\equiv 
	\exp\bigl[-(H_{\l}-\mathcal{F}_{\Lb})/T\bigr]
\end{equation}
and $\mathcal{F}_{\Lb}$ denoting the Helmholtz free energy.
The generalized forces \eqref{SThDForces} can then be expressed as 
\begin{equation}\label{SAdForces}
f^\mu_t = \mathcal{F}^\mu_{\Lb\ix{t}} \quad\text{with}\quad
	\mathcal{F}^\mu_{\Lb}\equiv \tr{\varrho_{\Lb} F_{\Lb}^\mu}
	=-\partial_\mu \mathcal{F}_{\Lb},
\end{equation}
where we have introduced the force operators 
\begin{equation}\label{AdForceOp}
F^w_{\Lb}\equiv - \partial_{\l} H_{\l} \quad\text{and}\quad
	F^u_{\Lb}\equiv - \ln\varrho_{\Lb}
\end{equation}
for later purposes. 
Inserting \eqref{SAdForces} into \eqref{SDissEnergy} shows that the
energy loss $A$ goes to zero in the quasi-static limit; 
the efficiency \eqref{SEfficiency} thus attains its upper bound $1$.
However, since the condition \eqref{SAdState} can be met only for 
infinitely long cycle times, the generated power, $P\equiv W/\T$,
also vanishes and the engine becomes virtually useless. 

Increasing the driving speed leads to finite power but inevitably 
also to energy losses reducing $\eta$. 
This trade-off can be understood quantitatively in the adiabatic 
response regime, where the external parameters change slowly compared
to the relaxation time of the system.
Under this condition, the thermodynamic forces \eqref{SThDForces} and 
the control rates $\dot{\Lb}_t$ are connected by the linear relations
\begin{equation}\label{ARKinEq}
f^\mu_t = \mathcal{F}^\mu_{\Lb\ix{t}} + R^{\mu\nu}_{\Lb\ix{t}}\dot{\L}^\nu_t,
\end{equation}
where $\nu=w,u$ and the adiabatic response coefficients 
$R^{\mu\nu}_{\Lb\ix{t}}$ depend parametrically on the driving 
protocols $\Lb_t$ \cite{Alessio2014}. 
The average energy loss \eqref{SDissEnergy} thus becomes 
\begin{equation}\label{ARLength}
A=\tint g^{\mu\nu}_{\Lb_t}\dot{\L}^\mu_t\dot{\L}^\nu_t
	\quad\text{with}\quad
	g^{\mu\nu}_{\Lb}\equiv -(R^{\mu\nu}_{\Lb}+ R^{\nu\mu}_{\Lb})/2
\end{equation}
denoting the elements of a, possibly degenerate, metric tensor in the
space of control parameters \footnote{Note that the matrix 
$g^{\mu\nu}_{\Lb}$ must be positive semi-definite, since the 
second law requires $A\geq 0$ for any closed path $\gamma$ and 
any parameterization.}. 
Thus, the Cauchy-Schwarz inequality implies
\begin{equation}\label{ARCauchySchwarz}
A\geq \LL^2/\T, \quad\text{where}\quad
	\LL\equiv\oint_\gamma\sqrt{g^{\mu\nu}_{\Lb}d\L^\mu d\L^\nu}
\end{equation}
corresponds to the thermodynamic length of the path $\gamma$.

Expanding the efficiency \eqref{SEfficiency} to second order in the 
driving rates $\dot{\Lb}_t$ yields $\eta = 1- A/\mathcal{W}$, where
the adiabatic work can be expressed as a line integral, 
\begin{equation}\label{ARWork}
\W= -\oint_\gamma \A^\mu_{\Lb}d\L^{\mu\strut}\quad\text{with}\quad
	\A^\mu_{\Lb} \equiv\partial_\mu	\mathcal{F}^{w}_{\Lb}\L^{w}
\end{equation}
being the thermodynamic vector potential.  
Upon using \eqref{ARCauchySchwarz}, we thus arrive at our first
main result, the power-efficiency trade-off relation
\begin{equation}\label{ARTradeOff}
(1-\eta)(\W/\LL)^2\geq \W/\T = P.
\end{equation} 
This bound implies that the power of any cyclic heat engine covered by
our model must vanish at least linearly as its efficiency approaches
the ideal value $1$. The maximal slope of this decay is determined by 
the thermodynamic mean force $\W/\LL$, where $\mathcal{L}$ and
$\mathcal{W}$ are geometric quantities, i.e., they are independent of 
the parameterization of the control path $\gamma$, see the lower panel
of Fig.~\ref{FigStirling}.

Moreover, \eqref{ARTradeOff} entails a universal optimization 
principle, which arises from the observation that the bound 
\eqref{ARCauchySchwarz} becomes an equality if the path $\gamma$ is
parameterized in terms of its thermodynamic length. 
To this end, $t$ has to be replaced with the speed function $\phi_t$,
which is implicitly defined through the relation
\begin{equation}\label{ARNatPara}
t=\T \int_0^{\phi_t} \!\!\! ds \sqrt{g^{\mu\nu}_{\Lb\ix{s}}
	\dot{\L}_{s}^\nu\dot{\L}_s^\mu }/\LL.
\end{equation}
Since $\W$ is not affected by this transformation, the bound
\eqref{ARTradeOff} can be saturated for any given control path 
$\gamma$ by choosing this optimal parameterization.
The efficiency then attains its geometric maximum 
\begin{equation}\label{ARMaxEff}
\eta^\ast= 1-\LL^2/\W\T. 
\end{equation}

Holding for any thermodynamically consistent microdynamics, our
general analysis so far applies to classical and quantum heat engines
alike.
To explore the fundamental differences between these two regimes, we 
now model the time evolution of the working medium explicitly using 
the well-established adiabatic Lindblad approach.  
This scheme rests on the assumption that the modulations of the
system Hamiltonian and the rate at which the external heat source 
provides thermal energy are both slow compared to the relaxation time
of the environment.
Applying this condition together with the standard weak-coupling 
approximation and a coarse-graining in time to wipe out memory effects
and fast oscillations yields the Markovian master equation
\begin{align}\label{CohLindblad}
&\partial_t\rho_t = \mathsf{L}_{\Lb\ix{t}}\rho_t \quad\text{with}\\
&\mathsf{L}_{\Lb}X\equiv -\frac{i}{\hbar}[H_\l,X] +\sum\nolimits_\sigma
\Bigl([V^\sigma_{\Lb}X,V^{\sigma\dagger}_{\Lb}] + 
	  [V^\sigma_{\Lb},X V^{\sigma\dagger}_{\Lb}]\Bigr).\nonumber
\end{align}
Here, $\hbar$ denotes Planck's constant and the Lindblad generator 
$\mathsf{L}_{\Lb}$ depends parametrically on the driving protocols
$\Lb_t$, for details see 
\cite{Alicki1979b,Albash2012,Majenz2013,Brandner2016}. 
Using \eqref{CohLindblad}, the periodic state $\rho_t$ can be 
determined by means of an adiabatic perturbation theory 
\cite{Cavina2017a,Weinberg2017}.

This procedure, which we outline in \cite{SM}, yields the 
Green-Kubo type expression 
\begin{equation}\label{CohAdResCoeff}
R^{\mu\nu}_{\Lb} = - \frac{1}{T} \int_0^\infty \!\!\! dt \;
	\bigl\llangle \exp[\mathsf{K}_{\Lb}t]F^\mu_{\Lb}
	\bigr|F^\nu_{\Lb}\bigr\rrangle
\end{equation} 
for the adiabatic response coefficients, where the canonical 
correlation function is defined as 
\begin{equation}\label{CohCanCorrF}
\llangle X | Y\rrangle \equiv \int_0^1\!\!\! dx \; 
	 \tr{\varrho_{\Lb}^{1-x}X\varrho_{\Lb}^x Y}
	-\tr{\varrho_{\Lb}X}\tr{\varrho_{\Lb}Y}
\end{equation}
for arbitrary observables $X$ and $Y$; the force operators 
$F^\mu_{\Lb}$ were introduced in \eqref{AdForceOp} and 
$\mathsf{K}_{\Lb}$ denotes the adjoint Lindblad generator, 
which is defined by the relation
$\tr{X\mathsf{K}_{\Lb}Y}\equiv\tr{Y\mathsf{L}_{\Lb} X}$
\footnote{Specifically, the adjoint generator reads
\begin{equation*}
\mathsf{K}_{\Lb} X\equiv \frac{i}{\hbar}[H_\l,X]
	+\sum\nolimits_\sigma \Bigl(
	V^{\sigma\dagger}_{\Lb}[X,V^\sigma_{\Lb}] +
	[V^{\sigma\dagger}_{\Lb},X]V^\sigma_{\Lb}\Bigr). 
\end{equation*}}.

This super operator is subject to three general consistency 
requirements. 
First, since we now work on a coarse-grained time scale, where
coherent oscillations have been averaged out, the operators 
$V^\sigma_{\Lb}$ can only induce jumps between the energy levels of 
the working system \cite{Majenz2013}. 
Hence, the eigenstates of $H_\l$ form the preferred basis of the 
dynamics and $\mathsf{K}_{{{\Lb}}}$ obeys the invariance condition
\begin{equation}\label{CohGenInvProp}
\mathsf{K}_{\Lb}[H_{\l},X] = [H_{\l},\mathsf{K}_{\Lb}X]. 
\end{equation}
Second, owing to microreversibility, the generators $\mathsf{K}_{\Lb}$
and $\mathsf{L}_{\Lb}$ are connected by symmetry relation 
\cite{Brandner2016}
\begin{equation}\label{ResRelQuDetBal}
\mathsf{T}\varrho_{\Lb}\mathsf{K}_{\Lb}X
	=\mathsf{L}_{\Lb}\varrho_{\Lb}\mathsf{T} X,
\end{equation}
where the super operator $\mathsf{T}$ induces time reversal
\cite{Brandner2016} and we assume that no magnetic field is applied to
the system, i.e., $\mathsf{T} H_{\l}=H_{\l}$ and 
$\mathsf{T}V^\sigma_{\Lb} =V^\sigma_{\Lb}$.
Together with \eqref{CohGenInvProp}, this property implies the 
adiabatic reciprocity relation $R^{\mu\nu}_{\Lb} = R^{\nu\mu}_{\Lb}$,
which resembles the familiar Onsager symmetry of linear
irreversible thermodynamics \cite{Onsager1931,Onsager1931a}.
Third, as a technical requirement, we understand that the jump 
operators $V^\sigma_{\Lb}$ form a self-adjoint and irreducible set; 
this condition ensures that, for $\Lb$ fixed, the mean of any 
observable relaxes to its unique equilibrium value under the dynamics
generated by $\mathsf{K}_{\Lb}$ in the Heisenberg picture
\cite{Spohn1977}. 
The expression \eqref{CohAdResCoeff} is then well-defined over the 
entire space of control parameters. 

We are now ready to analyze the impact of quantum effects on slowly
driven heat engines from a geometric perspective. 
To this end, we first divide the mechanical force operator into a 
diagonal and a coherent part, 
\begin{equation}
F^w_{\Lb} \equiv F^d_{\Lb} + i [H_{\l},G_{\l}]
	\equiv F^d_{\Lb} + F^c_{\Lb}.
\end{equation}
Here, $F^d_{\Lb}$ commutes with $H_{\l}$ and $G_{\l}$ corresponds to 
an adiabatic gauge potential \cite{Weinberg2017}. 
Upon inserting this decomposition into \eqref{CohAdResCoeff}, the 
adiabatic response coefficients decay into two components,
\begin{equation}\label{CohAdResCoeffDecomp}
R^{\mu\nu}_{\Lb} 
	= D^{\mu\nu}_{\Lb} + \delta_{\mu w}\delta_{\nu w} C_{\Lb}^{ww},
\end{equation}
where $D^{\mu\nu}_{\Lb}$ and $C_{\Lb}^{ww}$ are given by the formula 
\eqref{CohAdResCoeff} with $F^w_{\Lb}$ replaced by $F^d_{\Lb}$ and 
$F^c_{\Lb}$, respectively; the cross-terms between $F^c_{\Lb}$ and
the diagonal operators $F^d_{\Lb}$ and $F^u_{\Lb}$ vanish due to the
property \eqref{CohGenInvProp} of the adjoint generator. 
Next, by plugging \eqref{CohAdResCoeffDecomp} into the definition
\eqref{ARCauchySchwarz} of the thermodynamic length and using the 
concavity of the square-root function, we arrive at the bound
\footnote{To derive \eqref{CohTDLength}, observe  that 
$\mathcal{L}\geq \sqrt{\alpha}\mathcal{L}_d + \sqrt{1-\alpha}
\mathcal{L}_c$ for $0\leq \alpha\leq 1$ by the concavity of the 
square root function. 
Maximizing the right-hand side of this inequality with respect to 
$\alpha$ yields the desired result.}
\begin{equation}\label{CohTDLength}
\mathcal{L} \geq	\sqrt{\mathcal{L}^2_d+\mathcal{L}^2_c}. 
\end{equation}
The two quantities on the right, which are defined as
\begin{equation}
\nonumber
\mathcal{L}_d \equiv
	\oint_\gamma \sqrt{-D^{\mu\nu}_{\Lb}d\L^\mu d\L^\nu}
	\;\;\;\text{and}\;\;\;
	\mathcal{L}_c \equiv
	\oint_\gamma \sqrt{-C_{\Lb}^{ww} d\L^w d\L^w},
\end{equation}
thereby describe two genuinely different types of energy losses:
the reduced thermodynamic length $\mathcal{L}_d$ accounts for the 
dissipation of heat in the environment and the quantum correction
$\mathcal{L}_c$ arises from the decay of superpositions between the
energy levels of the working system, a mechanism known as quantum
friction \cite{Kosloff2002,Feldmann2003,Feldmann2004,Feldmann2006,
Feldmann2012}. 

The constraint \eqref{CohTDLength} puts an upper limit on the 
optimal finite-time efficiency \eqref{ARMaxEff}. 
This bound,
\begin{equation}\label{CohBndGeoForce}
\eta^\ast\leq 1-(\mathcal{L}_d^2+\mathcal{L}_c^2)/\mathcal{W}\tau, 
\end{equation}
which is our second main result, is saturated in the quasi-classical
limit, where $F^c_{\Lb}=0$; the energy eigenstates of the system are
then time-independent and the periodic state $\rho_t$  is diagonal in
this basis throughout the cycle.  
In fact, since the adiabatic work $\mathcal{W}$ is independent of
$F^c_{\Lb}$, the bound \eqref{CohBndGeoForce} shows that injecting
coherence into the working system can only reduce the maximum
efficiency of the engine at given power.
These coherence-induced performance losses are a universal feature of 
the slow-driving regime, where superpositions between different 
energy levels are irreversibly destroyed by the environment before 
their work content can be extracted through mechanical operations. 
While similar conclusions were drawn before for specific models
\cite{Kosloff2002,Feldmann2003,Feldmann2004,Feldmann2006,Feldmann2012}
and small driving amplitudes \cite{Brandner2016,Brandner2017a}, our 
new bound \eqref{CohBndGeoForce} applies to any heat engine that is 
covered by Lindblad dynamics and operated in adiabatic response. 
Thus, it further corroborates the emerging picture that quantum
effects can enhance the performance of thermal machines only far from 
equilibrium \cite{Uzdin2015,Uzdin2016b,Brandner2017a}.

We will now show how our general results can be applied to practical 
devices.
To this end, we consider a simple model for a solid-state quantum heat
engine that is inspired by a recent experiment \cite{Ronzani2018}. 
The working system consists of a superconducting qubit with 
Hamiltonian
\begin{equation}\label{ExHamiltonian}
H_{\l} = -\frac{\hbar\Omega}{2}\Bigl(\ce\sigma_x 
	+\sqrt{\l^2-\ce^2}\sigma_z\Bigr). 
\end{equation}
Here, $\sigma_x$ and $\sigma_z$ are the usual Pauli matrices, 
$\hbar\Omega$ denotes the overall energy scale and the dimensionless
parameters $\ce\geq 0$ and $\l\geq\ce$ correspond to the tunneling 
energy and the flux-tunable level-splitting of the qubit
\cite{Niskanen2007,Karimi2016a}. 
The role the environment is played by a normal-metal island, whose
temperature can be accurately controlled with established techniques
\cite{Giazotto2006} and monitored by means of sensitive electron 
thermometers, a technology that could soon enable calorimetric work
measurements \cite{Campisi2015,Viisanen2015,Gasparinetti2015,
Kupiainen2016a,Donvil2018a,Wang2018a}. 
This reservoir can be described in terms of two jump operators, 
$V^+_{\Lb}$ and $V^-_{\Lb}$, defined by the conditions 
\begin{align*}
&[H_{\l},V^\pm_{\Lb}] = \pm \hbar\Omega\l V^\pm_{\Lb}, \;\;\;
& \tr{V^{\pm\phantom{\dagger}}_{\Lb}\!\!V^{\pm\dagger}_{\Lb}}
	=\frac{\pm\Gamma\Omega\l}{1-\exp[\mp\hbar\Omega\l/T]},
\end{align*}
where $\Gamma$ determines the average jump frequency.  

We proceed in three steps. 
First, we evaluate the adiabatic response coefficients for the 
single-qubit engine using the formula \eqref{CohAdResCoeff}. 
Second, we calculate the geometric quantities entering the bounds 
\eqref{ARTradeOff} and \eqref{CohBndGeoForce} and the optimal 
speed function $\phi_t$ defined in \eqref{ARNatPara}. 
For simplicity, we thereby assume that the device is driven by 
harmonic temperature and energy modulations, i.e., we set
\begin{equation}\label{ExProt}
\Lb_t = \bigl(\hbar\Omega(1 + \sin^2[\pi\Omega t]),
	1+\sin^2[\pi\Omega t+\pi/4]\bigr).
\end{equation}
Hence, the control path $\gamma$ is a circle in the $\L^u-\L^w$ plane.
Third, in order to assess the quality of our bounds, we determine the 
periodic state $\rho_t$ of the system exactly by solving the 
time-inhomogeneous master equation \eqref{CohLindblad} for both 
constant and optimal driving speed.  
Using the expressions \eqref{SDefUW} and \eqref{SThDForces}, the power
and the efficiency of the engine can thus be obtained for any cycle 
time $\tau$. 

\begin{figure}
\includegraphics[scale=.45]{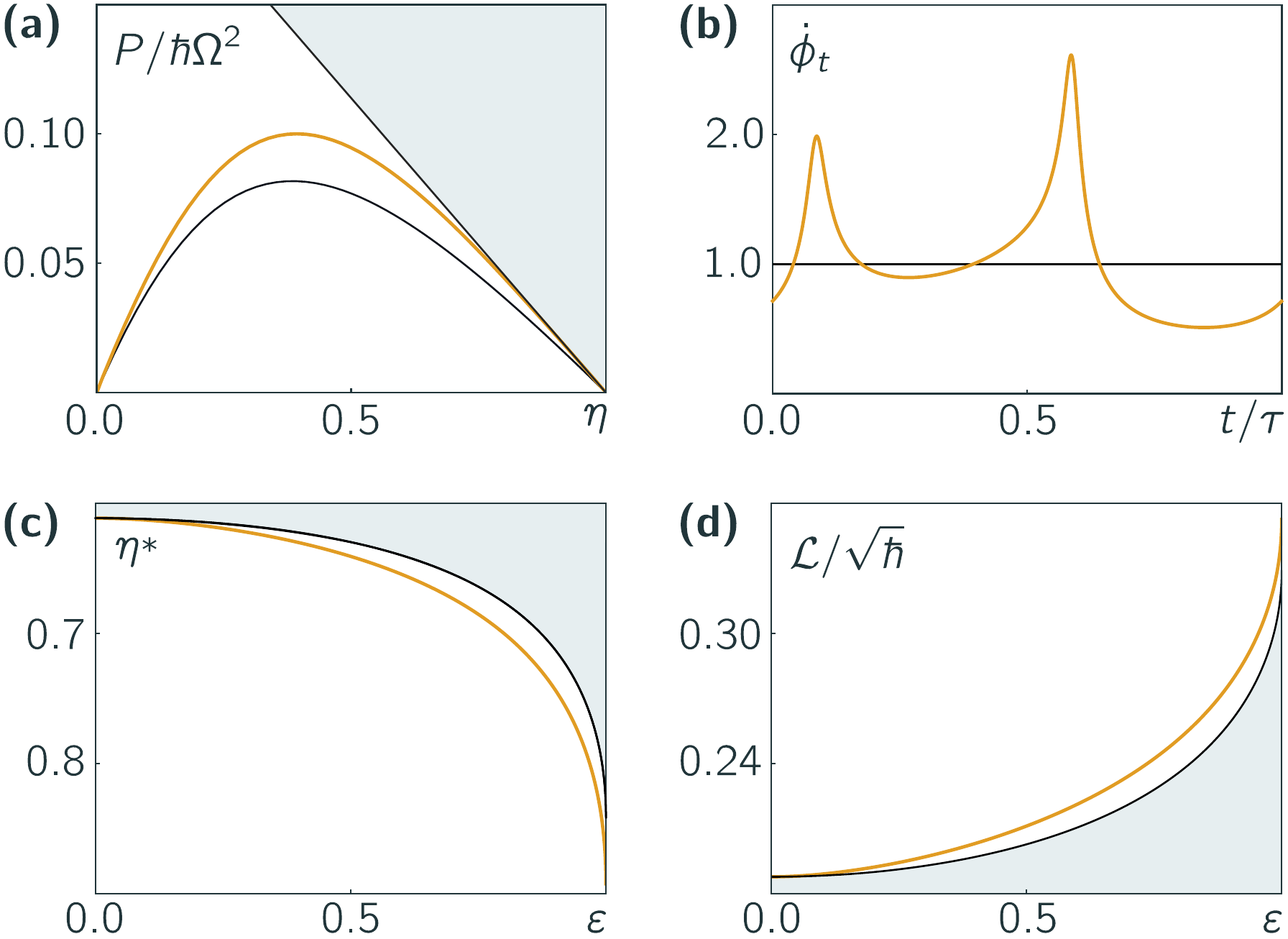}
\caption{Performance of a single-qubit engine. 
Upper panel: Geometric optimization. 
Plot (a) shows how the average power of the engine behaves compared to
its efficiency when the cycle time is varied from $\tau=1/10\Omega$ to
$\tau=50/\Omega$ for $\ce=3/5$.
The two curves are obtained for linear (black) and optimal (orange)
parameterization of the control path; the derivatives of the 
corresponding speed functions are shown in (b). 
The shaded area in (a) is inaccessible in adiabatic response by virtue
of our bound \eqref{ARTradeOff}.
Lower panel: Quantum losses. The orange curves show the optimal 
efficiency \eqref{ARMaxEff} for $\tau=3/\Omega$ (c) and the 
thermodynamic length (d) as a function of the coherence parameter
$\ce$; shaded areas indicate the bounds \eqref{CohBndGeoForce} and 
\eqref{CohTDLength}. For all plots, we have set $\Gamma=5$. 
\label{Fig_Engine}
\vspace*{-.6cm}
}
\end{figure}
The results of this analysis are summarized in Fig.~\ref{Fig_Engine},
for details see \cite{SM}. 
We find that, for optimal driving speed, our bound \eqref{ARTradeOff} 
is practically attained in the range $\eta\gtrsim 0.8$, which 
corresponds to $\tau\gtrsim 2/\Omega$.  
The optimal protocols $\Lb^\ast_t\equiv\Lb_{\phi\ix{t}}$ thereby 
outperform the harmonic profiles \eqref{ExProt} by roughly a factor
$1.2$ in power at given efficiency.
Remarkably, this increase in performance persists even for 
$\eta<0.8$, i.e., for short cycle times $\tau<2/\Omega$, which are not
covered by the slow-driving approximation \eqref{ARKinEq}. 
This phenomenon, whose degree of universality is yet to be 
established, raises the appealing perspective that it might be 
possible to extend our geometric description of microscopic heat 
engines beyond the limits of adiabatic response. 

The lower panel of Fig.~\ref{Fig_Engine} shows that the single-qubit 
engine operates most efficiently in the quasi-classical configuration
$\ce=0$. 
For this setting, the eigenstates of the Hamiltonian 
\eqref{ExHamiltonian} are independent of $\l$ and our bounds 
\eqref{CohTDLength} and \eqref{CohBndGeoForce} are saturated. 
Raising the value of $\ce$ leads to increasing quantum friction. 
Hence, the thermodynamic length grows and the optimal efficiency 
drops, whereby both figures closely follow their upper and lower 
bound, respectively. 
This behavior underlines our general result that coherence only
reduces the efficiency of thermodynamic cycles in adiabatic response.
Although this conclusion does not extend to the fast-driving regime, 
it still provides a valuable guideline for future theoretical and 
experimental studies seeking new strategies to gain a quantum 
advantage in the design of thermal machines. 

\newpage
\begin{acknowledgments}
K.B. thanks P. Menczel for insightful discussions and for a careful
proof reading of this manuscript and J. P. Pekola for helpful 
comments.
K.B. acknowledges support from Academy of Finland
(Contract No. 296073) and is associated with the Centre for Quantum 
Engineering at Aalto University.
K.S. was supported by JSPS Grants-in-Aid for Scientific Research 
(JP17K05587, JP16H02211).
\end{acknowledgments}

\end{document}



\title{Thermodynamic Geometry of Microscopic Heat Engines: Supplemental Material}
\maketitle

\vbadness=10000
\hbadness=10000

\makeatletter
\newcommand{\manuallabel}[2]{\def\@currentlabel{#2}\label{#1}}
\makeatother

\renewcommand{\H}{\mathsf{H}}
\renewcommand{\S}{\mathcal{S}}
\newcommand{\D}{\mathsf{D}}
\newcommand{\T}{\tau}
\newcommand{\F}{\mathcal{F}}
\newcommand{\R}{\mathcal{R}}
\newcommand{\W}{\mathcal{W}}
\newcommand{\A}{\mathcal{A}}
\renewcommand{\L}{\Lambda}
\newcommand{\LL}{\mathcal{L}}
\renewcommand{\r}{\rho}
\newcommand{\ce}{\varepsilon}
\renewcommand{\l}{\lambda}
\newcommand{\lb}{\boldsymbol{\l}}
\newcommand{\Lb}{\boldsymbol{\L}}

\newcommand{\tr}[1]{{{{\rm Tr}}}\bigl[ #1 \bigr]}
\newcommand{\av}[1]{\langle #1 \rangle}

\newcommand{\ixt}{\raisebox{-1.6pt}{{{\scriptsize $t$}}}}
\newcommand{\ix}[1]{\raisebox{-1.6pt}{{{\scriptsize $#1$}}}}

\renewcommand{\tint}{\int_0^\T \!\!\!\! dt \;}
\newcommand{\dbar}{{\mkern3mu\mathchar'26\mkern-12mu d}}

\newcommand{\sfL}{\mathsf{L}}
\newcommand{\sfK}{\mathsf{K}}
\newcommand{\sfH}{\mathsf{H}}
\newcommand{\sfD}{\mathsf{D}}
\newcommand{\sfT}{\mathsf{T}}

\manuallabel{MT_AdRespCoeff}{17}
\manuallabel{MT_CohInvCond}{19}
\manuallabel{MT_ARThState}{6}
\manuallabel{MT_STDForce}{3}
\manuallabel{MT_CohCanCorr}{18}
\manuallabel{MT_ARKinEq}{9}
\manuallabel{MT_ExHamiltonian}{25}
\manuallabel{MT_ARAdForces}{7}
\manuallabel{MT_ARForceOp}{8}
\manuallabel{MT_CohFroceOp}{21}
\manuallabel{MT_ExProt}{26}
\manuallabel{MT_Speed}{14}
\manuallabel{MT_SInOut}{2}
\manuallabel{MT_CohTSym}{20}
\manuallabel{MT_CohDecompAdRespCoeff}{22}

\newcommand{\ket}[1]{|#1\rangle}
\newcommand{\bra}[1]{\langle #1|}

\vbadness=10000
\hbadness=10000


\noindent
\section{Lindblad Dynamics and Adiabatic Response Theory}

In this section, we provide a brief technical review of the adiabatic
Lindblad approach to the dynamics of slowly driven open quantum 
systems, which is based on \cite{Brandner2016} and the references 
therein. 
We then derive the expression Eq.~\eqref{MT_AdRespCoeff} for the 
adiabatic response coefficients and discuss the general properties of
these quantities.

\subsection{Framework}
In the adiabatic Lindblad scheme, the time evolution of the state
$\rho_t$ of the working system is governed by the Markovian master 
equation 
\begin{equation}\label{LD_ME}
\partial_t \rho_t = \sfL_t\rho_t
	\equiv (-i\sfH_t + \sfD_t)\rho_t.
\end{equation}
Here, the two super operators 
\begin{align}
\label{LD_Gen}
\sfH X & \equiv [H,X]/\hbar
	\quad\text{and}\\
	\sfD X & \equiv \sum\nolimits_\sigma\Bigl(
	[V^\sigma X,V^{\sigma\dagger}] 
	+ [V^\sigma,X V^{\sigma\dagger}]\Bigr)\nonumber
\end{align}
describe the unitary dynamics of the bare system and the influence of
the environment, respectively. 
Note that, for convenience, we notationally suppress the dependence of
super operators, operators and scalar quantities on the control
parameters $\Lb$ throughout this section;
to indicate explicit dependence on the driving protocols $\Lb_t$, we
use the abbreviations $\mathsf{X}_{\Lb\ix{t}} \equiv\mathsf{X}_t$ and
$X_{\Lb\ix{t}}\equiv X_t$.

Owing to micro-reversibility, the dissipation super operator 
introduced in \eqref{LD_Gen} has to obey the quantum detailed balance 
condition
\begin{equation}\label{LD_DB}
\varrho\sfD^\ddagger X = \sfD \varrho X. 
\end{equation}
Here, $\varrho$ denotes the thermal state defined in 
Eq.~\eqref{MT_ARThState} and the adjoint of $\sfD$ with respect to 
the Hilbert-Schmidt scalar product is given by 
\begin{equation}
\sfD^\ddagger X \equiv \sum\nolimits_\sigma \Bigl(
	V^{\sigma\dagger}[X,V^\sigma]
	+[V^{\sigma\dagger},X]V^\sigma\Bigr).
\end{equation}
This relation has two important consequences. 
First, it implies the commutation rules
\begin{equation}\label{LD_ComRules}
\sfH\sfD= \sfD\sfH
	\quad\text{and}\quad
	\sfH\sfD^\ddagger= \sfD^\ddagger\sfH,  
\end{equation}
where the second identity is equivalent to the invariance 
condition Eq.~\eqref{MT_CohInvCond}, since $\mathsf{K}= 
i\sfH + \sfD^\ddagger$. 
Second, as a result of \eqref{LD_DB}, the two super operators 
$\sfH$ and $\sfD$ are both self-adjoint with respect to the
scalar product
\begin{equation}\label{LD_SP}
\langle X| Y\rangle \equiv 
	\int_0^1\!\!\! dx \;\tr{\varrho^{1-x}X^\dagger\varrho^x Y}
\end{equation}
in the operator space of the working system. 
Hence, there exists a joint set of normalized eigenvectors $M^j$
satisfying
\begin{equation}
\sfH M^j = \omega^j M^j, \quad
	\sfD^\ddagger M^j = \kappa^j M^j, \quad
	\langle M^j|M^k\rangle = \delta_{jk}, \nonumber
\end{equation}
where the eigenvalues $\omega^j$ and $\kappa^j$ are real and 
$\kappa^j\leq 0$.
Moreover, if the set of jump operators $V^\sigma$ is self-adjoint and
irreducible, we have $\kappa^0=0$ and 
$\kappa^j<0$ for $j>0$, where $M^0=\mathbbm{1}$ is the unique 
stationary eigenvector of $\sfD^\ddagger$.

\subsection{Adiabatic Response Coefficients}
The formula Eq.~\eqref{MT_AdRespCoeff} can now be derived by solving the
master equation \eqref{LD_ME} in the slow-driving regime using 
adiabatic perturbation theory \cite{Cavina2017a,Weinberg2017}. 
To this end, we apply the ansatz 
\begin{equation}\label{AGK_Ansatz}
\rho_t = \varrho_t + \int_0^1 \!\!\! dx \; 
	\varrho^{1-x}_t\xi_t\varrho^x_t,
\end{equation}
where $\xi_t$ is understood to be a small perturbation of first order
in the driving rates $\dot{\L}_t$. 
Inserting \eqref{AGK_Ansatz} into \eqref{LD_ME}, using the relations 
\eqref{LD_DB} and \eqref{LD_ComRules} and neglecting higher-order 
corrections yields the algebraic equation
\begin{equation}\label{AGK_AdCond}
(-i\sfH_t +\sfD_t^\ddagger)\xi_t 
	= (F^\mu_t -\langle\mathbbm{1}|F^\mu_t\rangle) 
	\dot{\L}^\mu_t/T_t.
\end{equation}
Since we assume that $M^0=\mathbbm{1}$ is the only stationary 
eigenvector of $\sfD^\ddagger$, the operator on the right of 
\eqref{AGK_AdCond} is orthogonal to the nullspace of the super 
operator acting on $\xi_t$ on the left. 
Hence, the unique solution of \eqref{AGK_AdCond} reads
\begin{equation}\label{AGK_PertState}
\xi_t = -\frac{1}{T}\int_0^\infty \!\!\! ds \; 
	\exp\bigl[(-i\sfH_t +\sfD^\ddagger_t)s\bigr]
	(F^\mu_t -\langle\mathbbm{1}|F^\mu_t\rangle)\dot{\L}^\mu_t. 
\end{equation}
Upon inserting this result into \eqref{AGK_Ansatz}, using the
definition Eq.~\eqref{MT_STDForce} of the thermodynamic forces and 
neglecting higher-order terms in the driving rates, we arrive at
\begin{align}
f^\mu_t - \langle\mathbbm{1}|F^\mu_t\rangle
&=	-\frac{1}{T}\int_0^\infty\!\!\! ds \; 
	\bigl\langle	(F^\nu_t -\langle\mathbbm{1}|F^\nu_t\rangle)\bigr|
	\exp[\mathsf{K}_t s]F^\mu_t\bigr\rangle\dot{\L}^\nu_t
	\nonumber\\
&=	-\frac{1}{T}\int_0^\infty\!\!\! ds \; 
	\bigl\langle \exp[\mathsf{K}_t s]F^\mu_t\bigr|
	(F^\nu_t -\langle\mathbbm{1}|F^\nu_t\rangle)\bigr\rangle\dot{\L}^\nu_t
	\nonumber\\
&=	-\frac{1}{T}\int_0^\infty\!\!\! ds \; 
	\bigl\llangle \exp[\mathsf{K}_t s]F^\mu_t\bigr|F^\nu_t\bigr\rrangle
	\dot{\L}^\nu_t.
\label{AGK_Derv}
\end{align}
Here, we have used that $\mathsf{K}=i\sfH +\sfD^\ddagger$ in the first
line; 
the second line follows by noting $\langle X | Y\rangle
=\langle Y|X\rangle^\ast =\langle Y^\dagger|X^\dagger\rangle$ and
$(\sfK X)^\dagger = \sfK X^\dagger$;
in the last line, we have inserted the definition 
Eq.~\eqref{MT_CohCanCorr} of the canonical correlation function. 
Finally, observing that $\langle\mathbbm{1}|F^\mu_t\rangle =
\mathcal{F}^\mu_t$ and comparing the last line of
\eqref{AGK_Derv} with Eq.~\eqref{MT_ARKinEq} yields the result 
Eq.~\eqref{MT_AdRespCoeff}. 

We now observing that Eq.~\eqref{MT_AdRespCoeff} can be rewritten as 
\begin{align}
\label{AGK_ProdForm}
R^{\mu\nu} &= -\frac{1}{T}\int_0^\infty\!\!\! dt\;
	\bigl\langle\exp[\mathsf{K}t]
	\delta F^\mu \bigr|\delta F^\nu \bigr\rangle
	\quad\text{with}\\
\delta F^\mu &\equiv F^\mu - \langle\mathbbm{1}|F^\mu\rangle. 
\nonumber
\end{align}
Using this expression, the three key properties of the coefficients 
Eq.~\eqref{MT_AdRespCoeff}, i.e., the positivity of the metric
$g^{\mu\nu} =-(R^{\mu\nu}+R^{\nu\mu})/2$, the reciprocity relation
$R^{\mu\nu}=R^{\nu\mu}$ and the decomposition law 
Eq.~\eqref{MT_CohDecompAdRespCoeff}, can be derived in a 
straightforward manner as we show in the following. 

\emph{1 - Positivity.} For arbitrary real variables $h_\mu$, we have
\begin{align}
\label{AGK_Pos}
T g^{\mu\nu}h_\mu h_\nu & = -\int_0^\infty\!\!\! dt\;
	\bigl\langle\exp[\sfK t]
	\delta F^h \bigr|\delta F^h \bigr\rangle\\
	& = -\int_0^\infty\!\!\! dt\;
	\bigl\langle	\delta F^h \bigr|
	\bigl(\exp[\sfK t]+\exp[\tilde{\sfK}t]\bigr)
	\delta F^h\bigr\rangle\bigr/2\nonumber\\
	& = -\sum\nolimits_{j>0}\kappa^j 
	|\langle M^j|\delta F^h\rangle|^2\bigl/|\kappa^j+i \omega^j|^2\geq 0. 
	\nonumber
\end{align}
Here, we have defined $\delta F^h\equiv h_\mu \delta F^\mu$ in the 
first line; the second line is obtained by observing 
\begin{align}
\langle \sfK X| X\rangle & = \langle X|\tilde{\sfK }X \rangle 
	\quad\text{and}\\
\langle \sfK X| X\rangle & = \langle X |\sfK X\rangle^\ast 
	= \langle X^\dagger | (\sfK X)^\dagger\rangle 
	= \langle X^\dagger | \sfK X^\dagger\rangle,\nonumber
\end{align}
where $\tilde{\sfK}\equiv - i\sfH + \sfD^\ddagger$; the third line in 
\eqref{AGK_Pos} follows by expanding $\delta F^h$ in the joint
eigenvectors of $\sfH$ and $\sfD$. 

\emph{2 - Reciprocity Relations.} For systems without a magnetic 
field, the generators $\sfK$ and $\sfL$ obey the symmetry relation 
Eq.~\eqref{MT_CohTSym}, which is equivalent to 
\begin{equation}
\sfT\sfK = \tilde{\sfK}\sfT
\end{equation}
by virtue of \eqref{LD_DB}. 
Here, $\sfT X=\Theta X \Theta^{-1}$ and $\Theta$ denotes the 
anti-unitary time-reversal operator. 
Hence, we have
\begin{equation}\nonumber
\langle \sfK \sfT X | \sfT Y\rangle 
	=\langle \sfT X| \tilde{\sfK}\sfT Y\rangle
	=\langle X^\dagger | \sfT^{-1}\tilde{\sfK}\sfT Y^\dagger\rangle
	=\langle X^\dagger | \sfK Y^\dagger\rangle. 
\end{equation}
Applying this identity to \eqref{AGK_ProdForm} and noting that 
$\sfT F^\mu = F^\mu$ for time-reversal symmetric systems yields the 
reciprocity relation $R^{\mu\nu} = R^{\nu\mu}$. 

\emph{3 - Decomposition.} Note that Eq.~\eqref{MT_CohFroceOp} is 
equivalent to 
\begin{equation}
\delta F^w = \delta F^d + i\sfH G = \delta F^d + F^c. 
\end{equation}
Inserting this line into \eqref{AGK_ProdForm}, recalling that 
$\sfH$ commutes with $\sfK$ and that $\sfH \delta F^d = 
\sfH \delta F^u =0$ yields the decomposition law 
Eq.~\eqref{MT_CohDecompAdRespCoeff}.

\section{Single-Qubit Heat Engine}
We provide further details on the example discussed in the last part
of the main text. 
Specifically, we first show how to determine the
geometric properties of this system and then explain how 
its finite-time thermodynamics can be analyzed beyond the adiabatic
response regime.

\subsection{Geometric Quantities}
We first observe that the Hamiltonian Eq.~\eqref{MT_ExHamiltonian} can
be decomposed as 
\begin{equation}
H_\l =\frac{\hbar\Omega\l}{2}\bigl(
	\ket{E^+_{\l}}\bra{E^+_{\l}}-\ket{E^-_{\l}}\bra{E^-_{\l}}\bigr),
\end{equation}
where the normalized eigenvectors are given by 
\begin{align}
& \ket{E^+_{\l}} = 
	\left(\!\begin{array}{r}
	\sin[\theta_{\l}/2]\\-\cos[\theta_{\l}/2]
	\end{array}\!\right)
	\quad\text{and}\quad
\ket{E^-_{\l}} = 
	\left(\!\begin{array}{r}
	\cos[\theta_{\l}/2]\\ \sin[\theta_{\l}/2]	
	\end{array}\!\right)
	\quad\text{with}\nonumber\\[9pt]
	& \sin[\theta_{\l}/2]\equiv \ce\Bigl/
	\sqrt{2\l^2+ 2\l\sqrt{\l^2-\ce^2}}.
\end{align}
Furthermore, the free energy and the adiabatic forces 
Eq.~\eqref{MT_ARAdForces} 
for the qubit are given by 
\begin{align}
\mathcal{F}_{\Lb} & = -T\ln\bigl[2\cosh[\hbar\Omega\l/2T]\bigr],\\
\mathcal{F}^u_{\Lb} & 
	=-\mathcal{F}_{\Lb}/T - (\hbar\Omega\l/2T)\tanh[\hbar\Omega\l/2T]
	\quad\text{and}\nonumber\\
\mathcal{F}^w_{\Lb} & 
	= (\hbar\Omega/2)\tanh[\hbar\Omega\l/2T].\nonumber
\end{align}
Using this result, the geometric work for an arbitrary engine cycle 
becomes
\begin{equation}
\mathcal{W}= \tint \mathcal{F}^w_{\Lb\ix{t}}\dot{\L}^w_t 
	= \frac{\hbar\Omega}{2}\tint \tanh[\hbar\Omega\l_t/2T_t] 
	\dot{\l}_t. 
\end{equation}
Note that this expression is independent of the coherence parameter 
$\ce$. 

For the remaining geometric quantities, we have to evaluate the 
adiabatic response coefficients Eq.~\eqref{MT_AdRespCoeff}. 
To this end, we proceed in three steps. 
First, we note that the jump operators defined in the main text are 
given by 
\begin{equation}
V^\pm_{\Lb} = \sqrt{\frac{\pm\Gamma\Omega\l}{
	1-\exp[\mp\hbar\Omega\l/T]}}\bigl(\Pi^x_\l \pm i\Pi^y_\l\bigr)
	\bigr/2,
\end{equation}
where 
\begin{align}
\Pi^x_\l&\equiv\ket{E^+_\l}\bra{E^-_\l}+\ket{E^-_\l}\bra{E^+_\l},\quad
\Pi^y_\l   \equiv i\ket{E^-_\l}\bra{E^+_\l}- i\ket{E^+_\l}\bra{E^-_\l},
	\nonumber\\
\Pi^z_\l &\equiv \ket{E^+_\l}\bra{E^+_\l} - \ket{E^-_\l}\bra{E^-_\l}. 
	\nonumber
\end{align}
The super operators $\sfD^\ddagger_{\Lb}$ and $\sfH_{\l}$ thus share
the normalized eigenvectors 
\begin{align}
M^0_{\Lb} =& \mathbbm{1},\\
M^z_{\Lb} =&\cosh[\hbar\Omega\l/2T]\Pi^z_\l
	+\sinh[\hbar\Omega\l/2T]\mathbbm{1}
	\quad\text{and}\nonumber\\
M^\pm_{\Lb} =& \sqrt{\frac{\hbar\Omega\l}{T\tanh[\hbar\Omega\l/2T]}}
	\bigl(\Pi^x_\l \pm i\Pi^y_\l\bigr)\bigr/2\nonumber
\end{align}
with corresponding eigenvalues 
\begin{align}
\label{SQE_SUEigenval}
& \kappa^0_{\Lb} = 0, 
	\quad \kappa^z_{\Lb} = -2\Gamma\Omega\l\coth[\hbar\Omega\l/2T],
	\quad \kappa^\pm_{\Lb} = \kappa^z_{\Lb}/2,\nonumber\\
& \omega^0_{\l} = \omega^z_{\l} =0, 
     \quad \omega^\pm_{\l} = \pm\Omega \l.
\end{align}
Second, we calculate the shifted force operators 
\begin{align}
\label{SQE_ShiftedFOp}
\delta F^u_{\Lb} &\equiv F^u_{\Lb} - \mathcal{F}^u_{\Lb}
	=\frac{\hbar\Omega\l}{2T\cosh[\hbar\Omega\l/2T]}M^z_{\Lb}
	\quad\text{and}\\
\delta F^w_{\Lb} &= \delta F^d_{\Lb} + F^c_{\Lb} \quad\text{with}\nonumber\\
\delta F^d_{\Lb} &\equiv F^d_{\Lb}-\mathcal{F}^w_{\Lb}\equiv 
	-\frac{\hbar\Omega}{2\cosh[\hbar\Omega\l/2T]}M^z_{\Lb}
	\quad\text{and}\nonumber\\[3pt]
F^c_{\Lb} & \equiv \frac{\ce\sqrt{\hbar\Omega\l T\tanh[\hbar\Omega\l/2T]}}{
	2\l\sqrt{\l^2-\ce^2}}\bigl(M^+_{\Lb} + M^-_{\Lb}\bigr),\nonumber
\end{align}
see Eq.~\eqref{MT_ARForceOp} and Eq.~\eqref{MT_CohFroceOp} for the 
general definitions of $F^\mu_{\Lb}$, $F^d_{\Lb}$ and $F^c_{\Lb}$. 
Third, upon inserting \eqref{SQE_ShiftedFOp} into \eqref{AGK_ProdForm}
and recalling that the operators \eqref{SQE_SUEigenval} are
eigenvectors of the generator $\mathsf{K}_{\Lb}$, we arrive at
\begin{equation}\label{SQE_ARCoeff}
\left(\!\begin{array}{cc}
R^{uu}_{\Lb} & R^{uw}_{\Lb}\\
R^{wu}_{\Lb} & R^{ww}_{\Lb} 
\end{array}\!\right)
	= \left(\!\begin{array}{cc}
	  -\l/T^3 & 1/T^2\\
	  1/T^2   & -1/\l T
	  \end{array}\!\right)\mathcal{R}_{\Lb}
	+ \left(\!\begin{array}{cc}
	 0 & 0\\ 0 &C^{ww}_{\Lb}
	\end{array}\!\right)
\end{equation}
and $D^{ww}_{\Lb}=R^{ww}_{\Lb}-C^{ww}_{\Lb}$ with
\begin{align}
C^{ww}_{\Lb} &= -\frac{\hbar\ce^2}{2\l^2(\l^2-\ce^2)}
	\frac{\Gamma}{\Gamma^2\coth^2[\hbar\Omega\l/2T]+1},\\
\mathcal{R}_{\Lb}&\equiv \frac{\hbar^2\Omega}{8\Gamma}
\frac{\tanh[\hbar\Omega\l/2T]}{\cosh^2[\hbar\Omega\l/2T]}.\nonumber
\end{align}

The quantities $\mathcal{L}$, $\mathcal{L}_d$ and $\mathcal{L}_d$ are
now obtained by inserting \eqref{SQE_ARCoeff} and the
protocols Eq.~\eqref{MT_ExProt} into the corresponding definitions
of the main text and carrying out the remaining integrals numerically.
To determine the optimal speed function $\phi_t$, we evaluate the 
expression
\begin{equation}
\phi^{-1}_{t\ix{k}} = \tau \int_0^{t\ix{k}}\!\!\! ds \; 
\sqrt{g^{\mu\nu}_{\Lb\ix{s}}\dot{\L}^\mu_s\dot{\L}^\nu_s}/\mathcal{L}, 
\end{equation}
which is equivalent to Eq.~\eqref{MT_Speed}, for a sufficiently large 
number of sampling points $0\leq t_k \leq\tau$ and determine the 
discretized speed function $\phi_{t\ix{k}}$ by point-wise inversion; 
the continuous function $\phi_t$ is recovered by interpolation.\\ 

\subsection{Exact Dynamics}
To analyze the performance of the single-qubit engine
without the adiabatic approximation Eq.~\eqref{MT_ARKinEq}, we 
have to solve the master equation 
\begin{equation}\label{SQE_ME}
\partial_t\rho_t = \mathsf{L}_{\Lb\ix{t}}\rho_t. 
\end{equation}
To this end, it is convenient to parameterized the periodic state 
$\rho_t$ in the form 
\begin{equation}
\rho_t \equiv \mathbbm{1}/2 + r^x_t \Pi^x_{\l\ix{t}} 
	+ r^y_t \Pi^y_{\l\ix{t}} + r^z_t \Pi^z_{\l\ix{t}}. 
\end{equation}
Plugging this ansatz in into \eqref{SQE_ME} yields the generalized
Bloch equations 
\begin{equation}\label{SQE_BlochEq}
\partial_t \left(\!\begin{array}{c}
	r^x_t \\ r^y_t \\ r^z_t
	\end{array}\!\right) = 
	\left(\!\begin{array}{ccc}
	-k^+_{\Lb\ix{t}} & -\Omega\l_t & -\theta'_{\l\ix{t}} \dot{\l}_t\\
	\Omega\l_t & -k^+_{\Lb\ix{t}} &0 \\
	\theta'_{\l\ix{t}}\dot{\l}_t  & 0 &-2k^+_{\Lb\ix{t}}
	\end{array}\!\right)
	\left(\!\begin{array}{c}
	r^x_t \\ r^y_t \\ r^z_t
	\end{array}\!\right) +
	\left(\! \begin{array}{c}
	0\\ 0\\ k^-_{\Lb\ix{t}}
	\end{array}\!\right)
\end{equation}
with
\begin{align}
k^\pm_{\Lb} \equiv \Gamma\Omega\l
	\frac{1\pm\exp[\hbar\Omega\l/T]}{\exp[\hbar\Omega\l/T]-1}
	\;\;\;\text{and}\;\;\;
	\theta'_{\l}\equiv \frac{\ce}{\l\sqrt{\l^2-\ce^2}}.
\end{align}
After inserting a specific set of driving protocols, i.e., for the 
setup discussed in the main text, either the harmonic profiles 
Eq.~\eqref{MT_ExProt} of the corresponding optimal protocols 
$\Lb_t^\ast =\Lb_{\phi\ix{t}}$, the differential equations 
\eqref{SQE_BlochEq} have to be solved numerically with respect to 
periodic boundary conditions. 
The exact input and output of the engine can then be determined by 
evaluating the expressions
\begin{align}
W & = -\hbar\Omega \int_0^\T \!\!\! dt \;\bigl( \l_t \theta'_{\l\ix{t}}r^x_t 
	+ r^z_t\bigr)\dot{\l}_t \quad\text{and}\\
U & = \frac{1}{2}\int_0^\T \!\!\! dt \; \bigl( 4r_t {{{\rm arctanh}}}[2r_t]
	+\ln[1/4-r^2_t]\bigr)\dot{T}_t, \nonumber
\end{align}
which follow directly from Eq.~\eqref{MT_SInOut}.
Note that we have used the short-hand notation 
$r_t \equiv \sqrt{(r^x_t)^2+(r^y_t)^2 + (r^z_t)^2}$ for the length of the 
Bloch vector.